\documentclass[conference]{IEEEtran}
\IEEEoverridecommandlockouts
\usepackage{cite}
\usepackage{amsmath,amssymb,amsfonts}
\usepackage{algorithmic}
\usepackage{graphicx}
\usepackage{textcomp}
\usepackage{xcolor}
\usepackage{enumitem}
\usepackage{multirow}
\def\BibTeX{{\rm B\kern-.05em{\sc i\kern-.025em b}\kern-.08em
    T\kern-.1667em\lower.7ex\hbox{E}\kern-.125emX}}
\begin{document}

\title{Blockchain Based Accounts Payable Platform for Goods Trade\\
}

\makeatletter
\newcommand{\linebreakand}{%
  \end{@IEEEauthorhalign}
  \hfill\mbox{}\par
  \mbox{}\hfill\begin{@IEEEauthorhalign}
}
\makeatother

\author{\IEEEauthorblockN{Krishnasuri Narayanam}
\IEEEauthorblockA{\textit{IBM Research, India} \\
knaraya3@in.ibm.com}
\and
\IEEEauthorblockN{Seep Goel}
\IEEEauthorblockA{\textit{IBM Research, India} \\
sgoel219@in.ibm.com}
\and
\IEEEauthorblockN{Abhishek Singh}
\IEEEauthorblockA{\textit{IBM Research, India} \\
abhishek.s@in.ibm.com}
\linebreakand
\IEEEauthorblockN{Yedendra Shrinivasan\IEEEauthorrefmark{1}}
\IEEEauthorblockA{\textit{Twitter, USA} \\
yshrinivasan@twitter.com}
\and
\IEEEauthorblockN{Parameswaram Selvam}
\IEEEauthorblockA{\textit{IBM Software Labs, India} \\
parselva@in.ibm.com}
\thanks{\IEEEauthorrefmark{1} The author contributed during his association with IBM}
}

\maketitle

\begin{abstract}
Goods trade is a supply chain transaction that involves shippers buying goods from suppliers and carriers providing goods transportation. Various business documents like purchase order, despatch advice, invoices, and receive advice get exchanged among the trade participants during any trade transaction. Similarly, various business processes like freight transport, invoice generation, goods receiving, invoice processing, and payment processing get executed by the participants in a trade transaction. Discrepancy during the execution of any of these processes leads to disputes between the participants involved, and the time consumed in resolving the disputes causes a delay in the process execution resulting in cost overhead for all the participants involved. Shippers are issued invoices from suppliers for the goods provided and from carriers for goods transportation. The shipper carries out goods receiving and invoice processing before proceeding to payment processing of bills for suppliers and carriers, where invoice processing includes tasks like processing claims and adjusting the payments. Goods receiving involves verification of received goods by Shipper's receiving team. Processing claims and adjusting the payments are carried out by Shipper's accounts payable team, which in turn is verified by the accounts receivable teams of suppliers and carriers. This paper presents a blockchain-based accounts payable system for shippers, which generates claims for deficiency in the goods received and accordingly adjusts the payment in the bills for suppliers and carriers. Primary motivations for these supply chain organizations to adopt blockchain-based accounts payable systems are to eliminate the process redundancies (accounts payable vs. accounts receivable), to reduce the number of disputes among the transacting participants, to reduce the dispute resolution time, and to accelerate the accounts payable processes via optimizations in the claims generation and blockchain-based dispute reconciliation.
\end{abstract}

\begin{IEEEkeywords}
Goods trade, accounts payable, invoice processing, dispute management, blockchain, smart contract
\end{IEEEkeywords}

\section{Introduction}
\label{sec:introduction}

\begin{figure*}[h]
	\centering
	\includegraphics[width=0.8\textwidth]{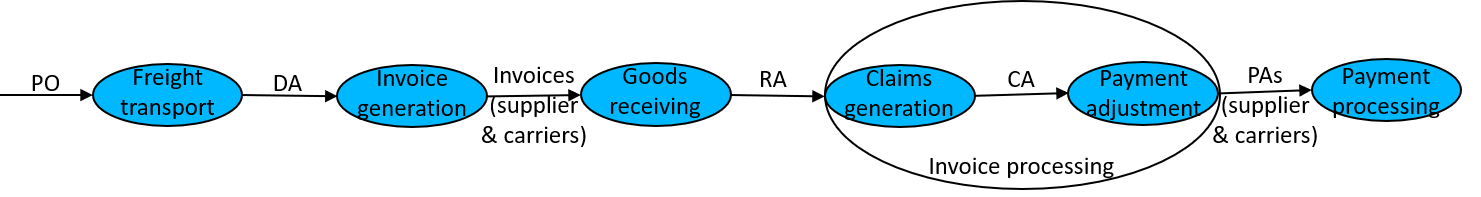}
	\caption{Various processes in goods trade (domestic or global)}
	\label{fig:goodsTradeProcesses}
\end{figure*}

\begin{figure*}[h]
    \centering
    \includegraphics[width=0.7\textwidth,height=8cm]{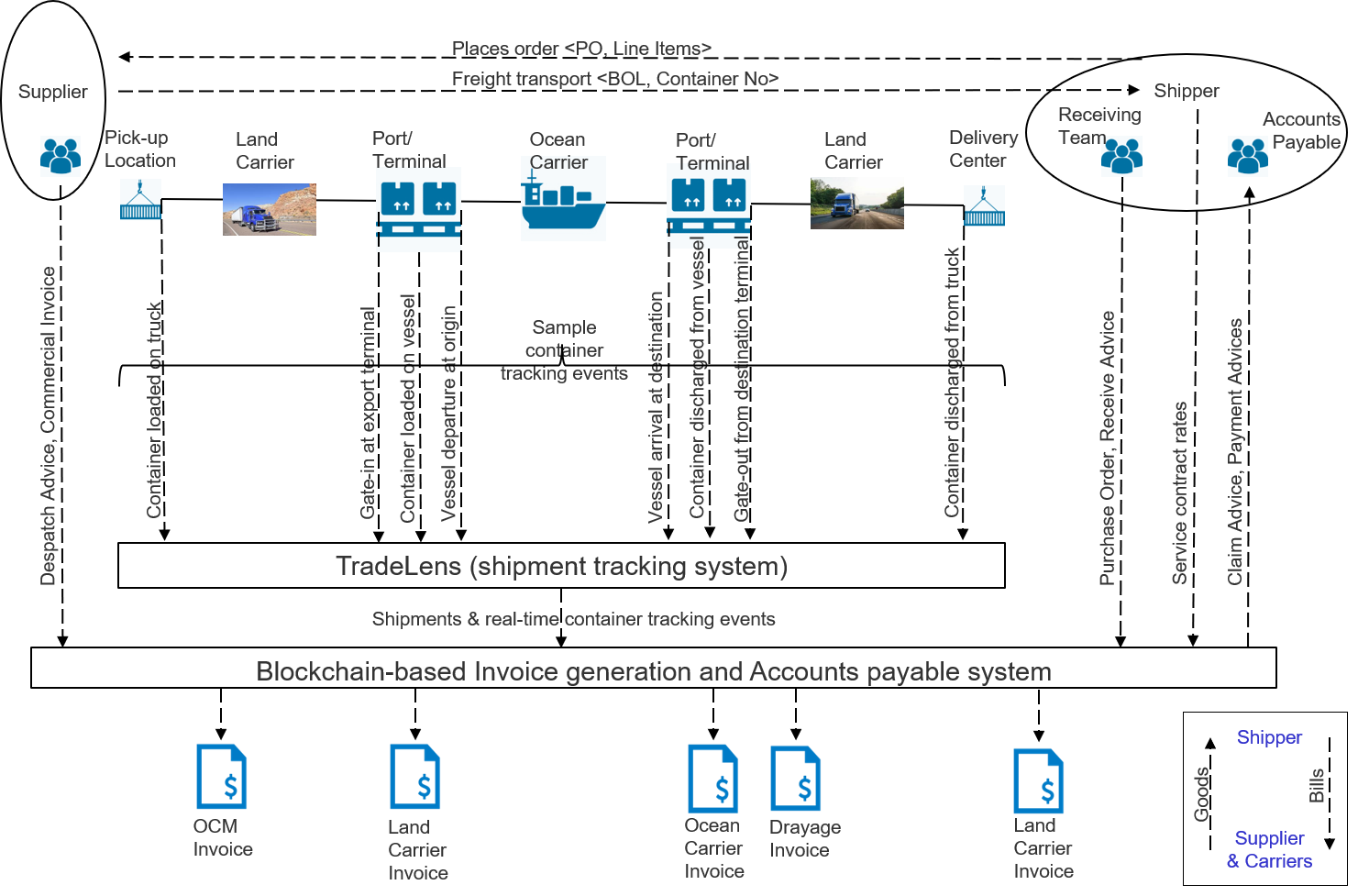}
    \caption{Blockchain-based global trade transaction processing}
    \label{fig:supplyChainProcesses}
\end{figure*}

Any trade transaction, be it domestic or global, involves exercising certain processes to complete. Domestic trade is the exchange of goods within country boundaries in contrast to between different countries in global/international trade.

We describe the different processes involved in goods trade using a global trade transaction in figure Fig. \ref{fig:supplyChainProcesses}.
% Need to list the different players part of a supply chain network ahead?
{\it Shippers} initiate a trade transaction by sending a {\it purchase order} (PO) which consists of details of the requested goods to the {\it suppliers}. Suppliers typically package the goods into intermodal containers either by themselves or with the help of {\it Origin Cargo Management (OCM)} team. Suppliers issue {\it despatch advice} (DA) that describes the goods packed details, and {\it commercial invoice} (CI) that describes the terms together with the details of the amount that shipper must pay for the goods supplied. Since global trade involves freight transportation across country borders, a typical freight journey involves multiple modes (e.g., road, rail, or sea) of carriers contributing to the container movement from origin to the destination. Moreover, freight transportation may also involve {\it drayage} providers to move containers a short distance via ground fright (e.g., move containers from truck to a ship).

Once freight reaches the delivery center at destination, the goods receiving team of the shipper verifies if the received goods can be accepted or not. If there are any damages to the received goods or discrepancies in terms of received quantity/price against PO, then the receiving team records the same via {\it receiving advice} (RA). The different carriers involved in freight movement also issue their respective invoices for their services. Once the shipper has access to invoices of carriers and the supplier, its accounts payable team needs to process the invoices. First, the accounts payable team raises a claim for the discrepancies reported in RA in the form of {\it claim advice} (CA). Second, the accounts payable team deducts the amount captured in CA from the appropriate invoice (either from the supplier's invoice or from a carrier's invoice whoever is accountable) and generate {\it payment advices} (PAs), where each PA captures the net amount payable by the shipper either to the supplier or to a carrier.

{\it Payment processing} involves executing a payment method as per the terms captured in the service contracts between trade participants. The most common payment method is {\it open account}, where the goods is shipped and delivered before the release of funds from the shipper within an agreed time frame. There is another payment method which involves {\it financing} facilitated by banks and financial institutions (e.g., {\it letter of credit} \cite{letterOfCredit} for supplier invoices in global trade, {\it factoring} \cite{factoring} for carrier invoices in global trade, and {\it reverse factoring} \cite{reverseFactoring} for supplier invoices in domestic trade). Goods trade can be summarized as the shipper acquiring the goods by paying the bills to the supplier and carriers as captured in figure Fig. \ref{fig:supplyChainProcesses}.

% please refer https://tex.stackexchange.com/questions/30985/displaying-a-wide-figure-in-a-two-column-document/30988

The sequence of processes that get exercised during goods trade is captured in figure Fig. \ref{fig:goodsTradeProcesses}. Different business documents are communicated among the transacting parties electronically in standard document formats known as {\it Electronic Data Interchange} (EDI) during goods trade. E.g., GS1 is one such EDI standard \cite{GS132}. There exist one or more separate freight invoices and one supplier invoice (i.e., CI) in the case of a global trade transaction. In contrast, supplier invoice typically includes freight charges as well in case of domestic trade.

The complexity of goods trade increases with the multiple en route handoffs between different parties involved in the goods movement process. End-to-end shipping visibility becomes significantly challenging to put together. Hence, shippers and other parties involved, try to gather as many details as possible during the execution of various goods trade processes (figure Fig. \ref{fig:goodsTradeProcesses}). Eventually, each organization ends up accruing data with their priorities and results in mistrust of information in the multi-party planning and invoice processing. To address the trust and transparency issues among the competitive and mutually distrusting participants of the supply chain network, blockchain-based the goods trade industry is adopting innovative solutions of late. TradeLens \cite{TLIBM} is a blockchain-based solution to provide visibility into the current status of {\it freight transport} with the help of real-time shipment tracking events originated from different supply chain participants. Generating the invoices for freight carriers (e.g., OCM, Land Carriers, Ocean Carriers, and Drayage) involved in the goods movement from origin to the destination using the real-time container tracking events and the shipment details captured by TradeLens as depicted in figure Fig. \ref{fig:supplyChainProcesses} got carried out in \cite{bitl2020}.
%Using TradeLens based real-time shipping events and shipment details, generating the invoices for carriers involved in the goods shipment process is carried out in \cite{bitl2020}.
In this paper, we propose a blockchain-based accounts payable system extending the TradeLens platform with capabilities to fulfill the needs of the supply chain network participants related to {\it invoice processing} (generation of CA and PAs) and dispute handling.

Here are the key features of our blockchain-based accounts payable platform. {\it CA} is generated by a blockchain smart contract using the EDI documents PO, DA, RA, and supplier invoice. {\it PAs} for supplier and carriers are generated by a blockchain smart contract using supplier invoice, carrier invoices, and CA. And our system allows the shipper, supplier, and carriers to raise disputes on the generated {\it claim advice} and {\it payment advices} and reconcile before {\it payment processing} with audit trails. Moreover, our system lets the reconciled and approved CA and PAs be sent to the customer's (shipper, supplier, or carrier) existing ERP systems via the API interfaces. Thus, our system is seamlessly integrated with the customer's downstream applications. Our system allows email/SMS {\it alert notifications} to be generated as per user notification settings based on different triggers (e.g., CA is issued or dispute raised).

In the following sections, first, we discuss related work in this domain and its shortcomings (Section \ref{sec:relatedWork}). Next, we present the architectural design of our system, describe various system components (Section \ref{sec:design}) and the implementation notes (Section \ref{sec:implementation}). Finally, we describe the experimental setup used for evaluation of the proposed system along with the performance of different types of transactions (Section \ref{sec:evaluation}) and summarize our contributions (Section \ref{sec:conclusion}).

\section{Related Work}
\label{sec:relatedWork}

Organizations are innovating products to help digitize various processes (Figure \ref{fig:goodsTradeProcesses}) involved in the goods trade industry. TradeLens \cite{TLIBM} is a solution to provide visibility into the current status of {\it freight transport} underpinned by blockchain technology. Producing blockchain-based e-invoices for the freight carriers as part of the {\it invoice generation} process got addressed in \cite{DLTLABS, bitl2020}.
%One related work in the domain of {\it invoice generation} is the blockchain-based freight and payment solution from the collaboration of Walmart Canada with DLT LABS \cite{DLTLABS}.
Several blockchain-based solutions exist related to financing during {\it payment processing} \cite{chang2020blockchain, chiu2019blockchain, bogucharskov2018adoption, wt}.
%\cite{chang2020blockchain, chiu2019blockchain, bogucharskov2018adoption, wt}.

There are instances where the use of blockchain technology for accounts payable (receivable) got discussed \cite{ipbyap, bcforap, bcinar}. Similarly, the possibility of carrying out matching (e.g., 3-way matching) of EDI documents on blockchain during {\it invoice processing} is also discussed \cite{bctoreplacesi}. However, all these proposals typically initiate {\it invoice processing} after the goods delivery to the shipper. The blockchain-based {\it invoice processing} system in our paper breaks down the {\it claims generation} into claims under different categories where claims under certain categories can be generated and issued before the goods delivery to the shipper. The advantages of this approach are two-fold: {\it invoice processing} gets accelerated since the dispute process can take place before the goods delivery to the shipper, and dispute reconciliation becomes easier since disputes are handled at the granularity of claims under a category resulting in faster dispute resolution. Hence, it is crucial to explore systems like the one proposed in this paper that enhance the effectiveness of blockchain-based accounts payable systems in {\it invoice processing}. Such a solution also helps organizations to build end-to-end blockchain-based platforms providing visibility across the various processes of goods trade.
\section{System Design}
\label{sec:design}
Figure \ref{fig:sysArch} represents the architecture of our proposed blockchain-based invoice processing system and the interaction among various system components.

%\subsection{Event Processor}
\label{sec:eventprocessor}
{\it Event Processor} (EP) component in the architecture obtains the details of shipments and their real-time tracking events from external shipment tracking systems like TradeLens. EP also provides an adaptor for receiving customized real-time shipment tracking events. All the events associated with a shipment are stored on the blockchain ledger. Only shipments of interest are managed, and their corresponding events are tracked to trigger {\it invoice processing} as the goods delivery to the shipper progresses.

%\subsection{Invoice Processor}
\label{sec:iacchaincode}
{\it Invoice Processor} (IP) is the chaincode component of our system. It implements various smart contract modules to simulate the {\it invoice processing} workflows. Smart contract module {\it computeCA} generates a CA for a given shipment using the EDI documents PO, DA, RA, and supplier invoice which are fetched from the blockchain ledger. Smart contract module {\it computePAs} generates PAs for the supplier and carriers using CA, invoices for supplier and carriers, and real-time shipment tracking events fetched from the ledger. Module {\it manageDispute} implements functions to create/raise a dispute, add comments with supporting documents to a dispute, update a dispute, resolve (accept/reject) a dispute on the CA/PAs generated by our system for each shipment. Module {\it finalizePA} finalizes a PA post which disputes cannot be allowed, while {\it autoApproveCAs} auto approve CAs with no disputes beyond a threshold number of days.

%\subsection{Goods Processor}
\label{sec:goodsprocessor}
{\it Goods Processor} (GP) component implements a set of REST APIs (with access control policies) for the user to interact with our {\it Accounts Payable} system and few cron jobs that invoke smart contract modules. GP component also maintains a PostgreSQL DB cache to boost the UI access performance. {\it EDI document access APIs} are used for accessing (send to \& retrieve from) EDI documents (PO, DA, RA, invoices for the supplier and carriers) corresponding to each shipment stored on the blockchain ledger. The same set of documents are hosted on Cloudant DB as well for enhancing the API access performance. {\it Query APIs} are used for querying shipments, generated CAs, and PAs from the blockchain ledger. {\it User management APIs} are used to manage the users of each participating organization in the blockchain network. {\it Notification APIs} are used to manage the subscription to the user alert notifications. {\it Status APIs} are used for increased user experience (e.g., transaction status monitoring and connecting with the client ERP systems). Cron job {\it generateClaimAdvices} invokes the smart contract function {\it computeCA} to generate the CA for each shipment. Cron job {\it generatePaymentAdvices} calls the smart contract module {\it computePAs} to generate PAs of each shipment, and cron job {\it CAautoApprove} calls the chaincode module {\it autoApproveCAs} to auto-approve CAs.

\begin{figure}
\centering
\includegraphics[scale=0.5]{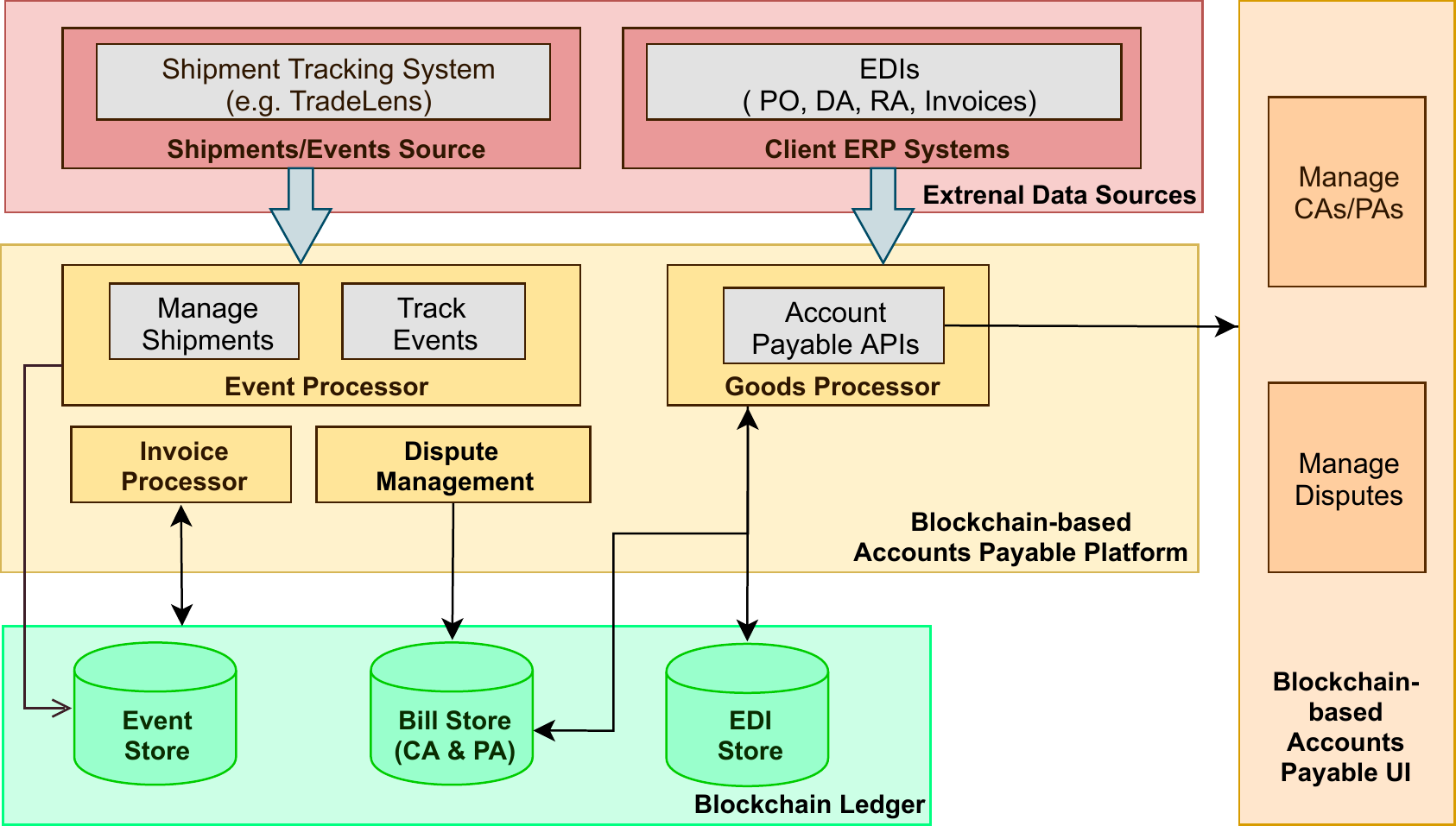}
%\caption{System Architecture and Transaction Flow}
\caption{Blockchain based invoice processing system architecture}
\label{fig:sysArch}
\end{figure}

%\subsection{Dispute Management}
\label{sec:disputemanagement}
{\it Dispute Management} component provides APIs (with access control policies) for the user to create/update a dispute, add comments on an existing dispute, and resolve a dispute (accept/reject). These APIs appropriately call the smart contract module {\it manageDispute} internally. This component also finalizes any PA with no active disputes by calling the smart contract module {\it finalizePA}.

%\subsection{Blockchain Ledger}
{\it Blochckain Ledger} component stores data related to {\it events} (real-time shipment tracking events), various EDI documents, and {\it bills} (generated CAs and PAs) shared among the blockchain network participants (e.g., shippers, suppliers, and carriers).

%\subsection{Blockchain Accounts Payable UI}
\label{sec:bitlui}
{\it Blockchain Accounts Payable UI} component provides an interface for users to interact with our system by calling the GP component REST APIs internally. The module {\it ManageCAs/PAs} provides an interface to view the computed CA and PAs for any shipment of interest to the user, while the module {\it ManageDisputes} provides an interface to raise and manage disputes on CA or PAs of a shipment.

%\subsection{External Data Sources}
{\it External Data Sources} component is a collection of external entities that interact with our system. The entity {\it Shipments/Events Source} can be any shipment tracking system for goods trade that generates real-time shipment tracking events (e.g., TradeLens). Shipments and events from this entity are consumed by the EP component. The entity {\it Client ERP Systems} feeds EDIs into our system and consumes the generated CA/PAs from our system.
\section{Implementation Details}
\label{sec:implementation}
The implementation details of {invoice processing} in our blockchain-based accounts payable system are presented here.

The notation used in this section is explained below.
\begin{itemize}[leftmargin=*]
\item {\it PO.Q:} Purchase order goods quantity.
\item {\it PO.P:} Price for each goods item as per purchase order.
\item {\it DA.Q:} Quantity dispatched by the supplier/quantity received by the shipper.
\item {\it RA.Q:} Quantity accepted by the shipper after the verification by shipper's goods receiving team.
\item {\it CI.Q:} Invoiced goods quantity by the supplier.
\item {\it CI.P:} Price for each goods item as per the supplier invoice.
\end{itemize}

\subsection{Claim Advice generation}
\label{sec:caComputation}
Smart contract module {\it computeCA} auto generates CA for each PO on behalf of the shipper's {\it accounts payable} team as goods delivery progresses by analyzing EDI documents like PO, DA, RA, and supplier invoice (i.e., CI in case of global trade). Claims generated on behalf of the accounts payable team compensate for any damages to the received goods along with discrepancies in terms of goods received quantity or price against the PO. The accounts payable team typically computes the claim amount by exercising a 2-way, a 3-way, or a 4-way matching \cite{payablesMatching, payablesMatching2, payablesMatching3} of the EDI documents. In a 2-way matching, the invoice is matched to the purchase order to validate the {\it two} criteria that $CI.Q \leq PO.Q$ and $CI.P \leq PO.P$. In a 3-way matching, the invoice is matched to the goods received to validate a {\it third} criteria that $CI.Q \leq DA.Q$. In a 4-way matching, the invoice is matched to the goods accepted to validate a {\it fourth} criteria that $CI.Q \leq RA.Q$.
%In 2-way matching, the invoice is matched to the purchase order to validate the following {\it two} criteria that the invoiced quantity ($CI.Q$) is less than or equal to the purchase order quantity ($PO.Q$), and the invoice price ($CI.P$) is less than or equal to the purchase order price ($PO.P$). In 3-way matching, the invoice is matched to the goods received to validate a {\it third} criteria that the invoiced quantity ($CI.Q$) is less than or equal to the received quantity ($DA.Q$). In 4-way matching, the invoice is matched to the goods accepted to validate a {\it fourth} criteria that the invoiced quantity ($CI.Q$) is less than or equal to the accepted quantity ($RA.Q$).

Matching EDI documents ensures that the shipper doesn't pay for goods that is not received or overpay for goods received. The shipper is entitled to pay to the supplier for the quantity of the accepted goods at the agreed price as per the purchase order, which is the amount $RA.Q * PO.P$. However, the payable amount as per the invoice by the shipper to the supplier is $CI.Q * CI.P$. The difference between the amount payable and the amount entitled to be paid is called as the claim amount to the shipper, which is equal to the amount $(CI.Q * CI.P) - (RA.Q * PO.P)$.

There can be price mismatch, or quantity mismatch, or both during the process of matching the EDI documents. We split the claims added by our blockchain-based accounts payable system to the CA document into {\it four} different claim categories.

\subsubsection{Claim categories under the scenario $CI.Q \leq PO.Q$}
\begin{figure}[ht]
    \centering
    \includegraphics[width=1.0\columnwidth]{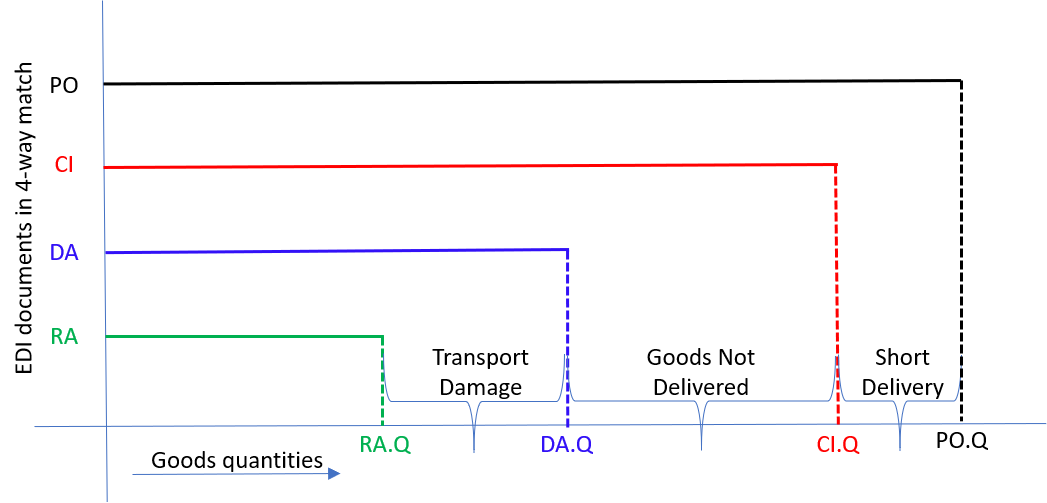}
    \caption{Goods quantity discrepancies in a 4-way match when CI.Q $\leq$ PO.Q}
    \label{fig:4wayMatchShort}
\end{figure}
Fig. \ref{fig:4wayMatchShort} depicts the possible goods quantity mismatch discrepancies during a {\it 4-way} match of the EDI documents PO, DA, RA, and CI under the scenario where $CI.Q \leq PO.Q$. The claims added to the CA document can be split into the following claim categories to represent the goods quantity mismatch, or price mismatch, or both.
\begin{itemize}[leftmargin=*]
\item {\it SHORT\_DELIVERY:} Difference in the quantity of goods between ordered (PO.Q) and invoiced (CI.Q) is captured as discrepancy under this category. $CI.Q \leq PO.Q$ implies partial fulfillment, which means that the supplier has not despatched all the goods quantity requested. The claim amount under this category is {\it 0}.
\item {\it PRICE\_DISCREPANCY:} Difference in the goods price between supplier invoiced (CI.P) and ordered by the shipper (PO.P) is captured as a discrepancy under this category. $(CI.P - PO.P)$ attributes to the price change from what is agreed. The claim amount under this category is $(CI.P-PO.P)*CI.Q$.
\item {\it GOODS\_NOT\_DELIVERED:} Difference in the quantity of the goods between supplier invoiced (CI.Q) and received by the shipper (DA.Q) is captured as discrepancy under this category. $CI.Q-DA.Q$ attributes to billed quantity mismatch. The claim amount under this category is $(CI.Q-DA.Q) * PO.P$.
\item {\it TRANSPORT\_DAMAGE:} Difference in the quantity of the goods between despatched by the supplier (DA.Q) and received by the shipper (RA.Q) is captured as discrepancy under this category. $DA.Q>RA.Q$ attributes to the shipper's goods receiving team not accepting the difference in the quantity $DA.Q-RA.Q$ as damaged goods. The claim amount under this category is $(DA.Q-RA.Q)*PO.P$.
\end{itemize}
The claim amount of a CA is the sum of the claim amounts under each claim category above. Thus the total claim amount of the CA is $ 0 + (CI.P-PO.P)*CI.Q + (CI.Q-DA.Q) * PO.P + (DA.Q-RA.Q)*PO.P$ which is equal to $(CI.Q * CI.P) - (RA.Q * PO.P)$.

\subsubsection{Claim categories under the scenario $CI.Q > PO.Q$}
\begin{figure}[ht]
    \centering
    \includegraphics[width=1.0\columnwidth]{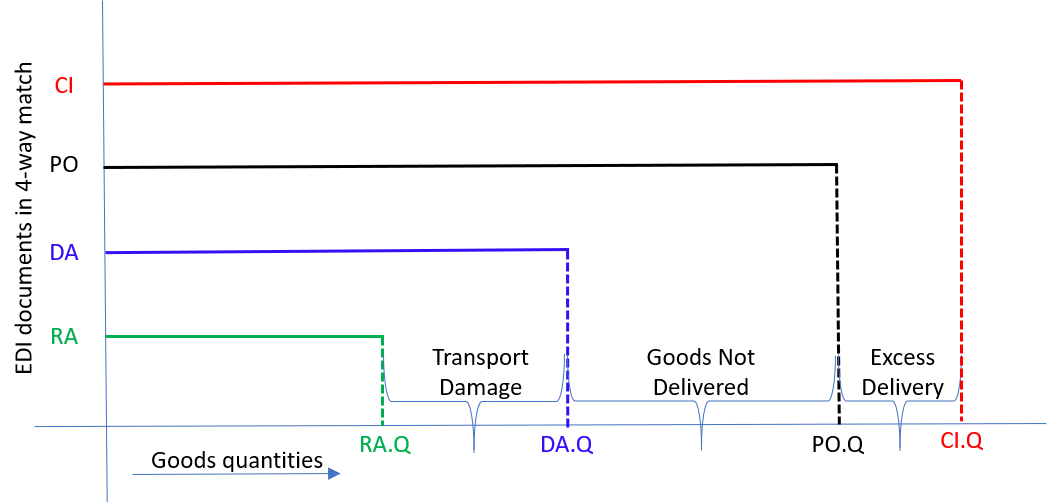}
    \caption{Goods quantity discrepancies in a 4-way match when $CI.Q > PO.Q$}
    \label{fig:4wayMatchExcess}
\end{figure}
Fig. \ref{fig:4wayMatchExcess} depicts the possible goods quantity mismatch discrepancies during a {\it 4-way} match of the EDI documents PO, DA, RA, and CI under the scenario where $CI.Q > PO.Q$. The claims added to the CA document can be split into the following claim categories to represent the goods quantity mismatch, or price mismatch, or both.
\begin{itemize}[leftmargin=*]
\item {\it EXCESS\_DELIVERY:} Difference in the quantity of goods between ordered (PO.Q) and invoiced (CI.Q) is captured as discrepancy under this category. $CI.Q > PO.Q$ implies that the supplier has despatched excess quantity than requested. The claim amount under this category is $(CI.Q-PO.Q)*PO.P$.
\item {\it PRICE\_DISCREPANCY:} Difference in the goods price between supplier invoiced (CI.P) and ordered by the shipper (PO.P) is captured as a discrepancy under this category. $(CI.P - PO.P)$ attributes to the price change from what is agreed. The claim amount under this category is $(CI.P-PO.P)*CI.Q$.
\item {\it GOODS\_NOT\_DELIVERED:} Difference in the quantity of the goods between supplier despatched (DA.Q) and ordered by the shipper (PO.Q) is captured as discrepancy under this category. $PO.Q-DA.Q$ attributes to billed quantity mismatch. The claim amount under this category is $(PO.Q-DA.Q) * PO.P$.
\item {\it TRANSPORT\_DAMAGE:} Difference in the quantity of the goods between despatched by the supplier (DA.Q) and received by the shipper (RA.Q) is captured as discrepancy under this category. $DA.Q>RA.Q$ attributes to the shipper's goods receiving team not accepting the difference in the quantity $DA.Q-RA.Q$ as damaged goods. The claim amount under this category is $(DA.Q-RA.Q)*PO.P$.
\end{itemize}
The claim amount of a CA is the sum of the claim amounts under each claim category above. Thus the total claim amount of the CA is $ (CI.Q-PO.Q)*PO.P + (CI.P-PO.P)*CI.Q + (PO.Q-DA.Q) * PO.P + (DA.Q-RA.Q)*PO.P$ which is equal to $(CI.Q * CI.P) - (RA.Q * PO.P)$.

Note that $RA.Q \leq DA.Q$ holds good under both the above scenarios $CI.Q \leq PO.Q$ and $CI.Q > PO.Q$ as depicted in Fig. \ref{fig:4wayMatchShort} and Fig. \ref{fig:4wayMatchExcess} respectively. And the total claim amount computed during the CA document generation under any scenario is the same and is equal to the amount $(CI.Q * CI.P) - (RA.Q * PO.P)$.

Moreover, by referring to Section \ref{sec:introduction}, the EDI documents PO, DA, and CI get ready for access on the supply chain network before the completion of the goods delivery to the shipper. And the EDI document RA gets ready for access only after the completion of the goods delivery to the shipper.

Hence our blockchain-based accounts payable system generates the CA in two passes:
\begin{itemize}[leftmargin=*]
\item {\it Pass-1 using only 3 EDI documents PO, DA \& CI:} Generate claims under the $PRICE\_DISCREPANCY$ and $GOODS\_NOT\_DELIVERED$ categories along with the category $SHORT\_DELIVERY$ (if $CI.Q \leq PO.Q$) or $EXCESS\_DELIVERY$ (if $CI.Q > PO.Q$) as applicable before the goods delivery to the shipper on the occurrence of event like {\it container loaded on truck} from TradeLens shipment tracking system (refer to figure Fig. \ref{fig:supplyChainProcesses}).
%Implements the CA generation tasks under the SHORT\_DELIVERY \& PRICE\_DISCREPANCY categories.
\item {\it Pass-2 using all 4 EDI documents PO, DA, CI \& RA:} Updates the CA generated in Pass-1 to add claims generated under the categories $TRANSPORT\_DAMAGE$ after the completion of goods delivery to the shipper post the occurrence of event like {\it container dispatched from truck} from TradeLens shipment tracking system (figure Fig. \ref{fig:supplyChainProcesses}).
\end{itemize}

This approach of breaking the CA generation into two passes in our blockchain-based accounts payable platform has an advantage with respect to letting the users (i.e., suppliers and carriers) access the CA much earlier than when the traditional accounts payable platforms would allow (i.e., where the users can access CA only after goods delivery to the shipper). Our system models the issuance of a CA and subsequent possible dispute (raised by supplier or carriers) workflows at the granularity of claims under each category.

\subsubsection{Additional details on Claim Amount in CA}
We assume that the claim amount can be negative without loss of generality, and a negative claim amount indicates the scenarios where the invoice is billed for fewer goods quantity than accepted by the shipper or billed at fewer goods price than agreed as per the purchase order among other possibilities. Additionally, whenever $DA.Q > PO.Q$, then we set $DA.Q = PO.Q$ as a preprocessing step. This ensures that the shipper never pays for goods quantity over the requested quantity $PO.Q$. And the shipper accepts $0\leq RA.Q \leq DA.Q$ from the entire goods dispatched by the supplier.
%Our system issues the claims under each task category as a sub claim. And disputes from the supplier and carriers can be raised on each sub claim independently. Once all the 4 types of sub-claims are available, the claim of a PO can be obtained by simply summing the sub claim amounts.

\subsection{Payment Advice generation}
\label{sec:paComputation}
The EP component extracts shipments from {\it TradeLens} and tracks them using the tuple $\langle bol, containerNo\rangle$ in our system, where {\it bol} corresponds to the carrier issued bill of lading for the shipment and {\it containerNo} corresponds to the multimodal container number into which the shipper's goods is packed. However, note that the goods against a PO may get shipped using multiple containers. Similarly, a container may ship goods belonging to multiple POs. Hence, a carrier invoice for each container is presented by splitting the invoice amount at $\langle PO, bol\rangle$ granularity by any freight carrier involved in the movement of the container.

Smart contract {\it computePAs} generates a PA for each supplier and carrier associated with a PO. PA for any freight carrier is generated by aggregating the carrier invoice amount (a given PO share only) for each container carrying the goods against a PO. PA for the supplier is generated by deducting the claim amount captured in CA (excluding the claim amount under TRANSPORT\_DAMAGE category) from the supplier invoice. The claim amount for damaged goods is deducted from the PA of either OCM (which is also considered as a carrier) or supplier, based on whoever had done the goods packing into the container (this information is captured by a customized real-time shipment tracking event in TradeLens shipment tracking system). The {\it computePAs} module can include any other custom method to distribute the incurred loss due to goods damage among the supplier and carriers.

\subsection{Claim Advice state management}
%\subsubsection{Claim Advice workflow}
The states of claim advice under any of the 4 possible claim categories in our system are determined based on our customers' interactions, which could be adjusted for different business processes. We implement the CA state transition in our system as a finite state machine. Figure \ref{fig:CAstates} depicts the allowed state transitions along with the criteria for each transition. Any CA (irrespective of its {\it claim category}) gets created with state {\it CIP (computation in progress)} as the goods delivery is under progress. Once the computation of claims under any category is complete, the CA under that category gets issued and moves to {\it Open} state. Any shipper, supplier, or carriers associated with the goods trade transaction can access the CA in {\it Open} state. If any of these supply chain participants find some discrepancy in the CA, they raise a dispute which needs to be reviewed by other participants hence CA goes into {\it awaiting resolution (AR)} state (e.g., if the shipper raises a dispute on CA then the supplier must review the dispute). Once the reviewer accepts/rejects the dispute on a CA, the CA moves to {\it manually approved (MA)} state. If a CA in {\it Open} state does not have any dispute within a waiting period (user-configurable), it is moved to {\it auto-approved (AA)} state by the cron job {\it CAautoApprove} of the GP component.
\begin{figure}[ht]
    \centering
    \includegraphics[width=1.0\columnwidth]{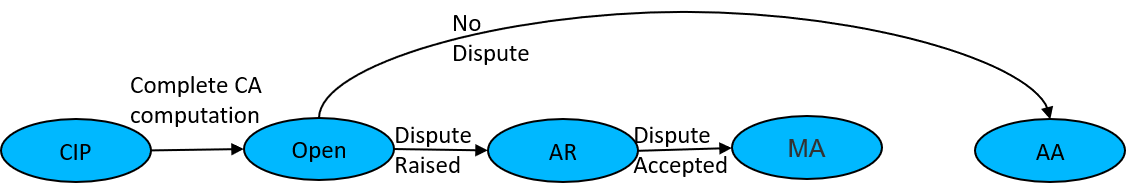}
    \caption{State transitions for claim advice (CA)}
    \label{fig:CAstates}
\end{figure}

\subsection{Payment Advice state management}
%\subsubsection{Payment Advice Workflow}
%We implement PA state transition in our system as a finite state machine. Figure \ref{fig:PAstates} depicts the allowed state transitions for PA along with the criteria for each transition. PA first gets created with state {\it computation in progress (CIP)} after the goods is delivered to the shipper. PA computation completes once CA is available for a PO, and PA moves into {\it awaiting resolution (AR)} state as it gets issued. The review process for PA is similar to CA. If there are no disputes, the shipper must finalize the PA following which the PA state moves to {\it manually approved (MA)}. There is no auto-approval support for PA.
We implement the PA state transition in our system as a finite state machine. Figure \ref{fig:PAstates} depicts the allowed state transitions for PA along with the criteria for each transition. PA first gets created with state {\it computation in progress (CIP)} after the goods delivery to the shipper. PA computation completes once CA is available for a PO, and PA moves into {\it awaiting resolution (AR)} state as it gets issued. The review process for PA is similar to CA. If there are no disputes, the shipper must finalize the PA following which it moves to {\it manually approved (MA)} state. There is no auto-approval support for PA.
\begin{figure}[ht]
    \centering
    \includegraphics[width=0.65\columnwidth]{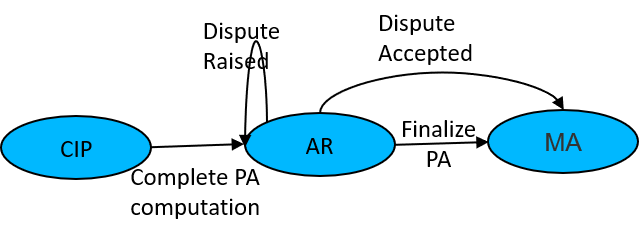}
    \caption{State transitions for payment advice (PA)}
    \label{fig:PAstates}
\end{figure}

\subsection{Claim Advice Dispute Workflow Comparison}
\begin{figure}[ht]
    \centering
    \includegraphics[width=1.0\columnwidth]{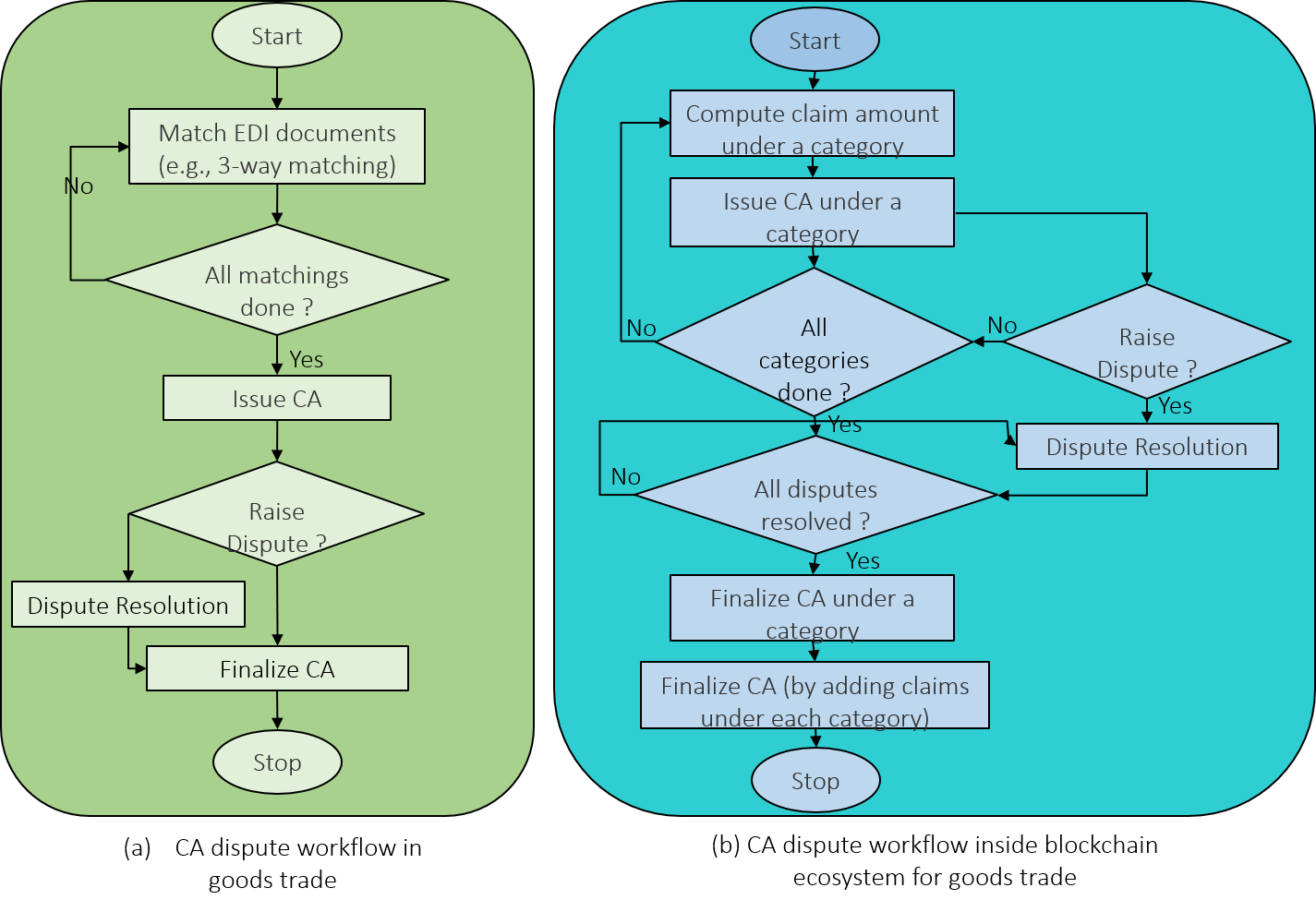}
    \caption{CA dispute workflow: traditional vs. blockchain-based}
    \label{fig:disputeWorkflow}
\end{figure}
Figure \ref{fig:disputeWorkflow} depicts the dispute workflow in traditional goods trade environments vs. the blockchain ecosystem for goods trade. In a traditional workflow, the dispute process starts after goods delivery to the shipper, and once the CA gets computed by the EDI documents matching and gets issued. In contrast, since our system computes the claim amount per category and issues claims under each category independently, a user can raise a dispute without the need to wait for the completion of goods delivery to the shipper. This pre-CA preview can be seen as a direct benefit of our blockchain-based ecosystem for goods trade. Another benefit is that, since all the information is available on the blockchain ledger and shared among all the network peers, CA reconciliation is much faster, resulting in quicker dispute resolution.

\subsection{Privacy Considerations}
Since the blockchain ecosystem for goods trade can host multiple shippers, suppliers, and carriers along with members from other organizations, a peer node in the blockchain network represents a member from any of these organizations. Data (EDIs, shipment tracking events, CA, and PAs) is shared among only a subset of the network peers to ensure data privacy. We exploit {\it channels} and {\it private data collections} mechanisms in the Hyperledger Fabric blockchain network for this purpose.  Moreover, we have capabilities built in our system to provide access to the shared data on the blockchain ledger based on the user role. Our system also allows the trade participants to update an existing smart contract in mutual agreement and seamlessly apply those changes retrospectively to all the shipments tracked in our system from the desired past date.

A shipper can place an order for multiple different items (called line items) in a single PO. Each PO line item may get supplied by a different supplier and get associated with the respective supplier invoice. Our system prevents access to CA and PA of other line items within the same PO from the supplier of a given line item by the following two means: (i) by generating CA and PA at the granularity of $\langle PO, lineItem\rangle$, and (ii) by appropriately configuring the privacy mechanisms and access policies discussed in this section.
\section{Evaluation}
\label{sec:evaluation}

%The aim of our experimental setup is to simulate the global trade ecosystem on blockchain platform. To achieve this, we map each participant of the ecosystem to a membership service provider (MSP) in the blockchain network. Each MSP consists of one or more peers of the blockchain network. The peers are deployed as docker containers running Hyperledger Fabric version $1.4.1$. Each peer is deployed on a dedicated SoftLayer ~\cite{softLayer} VM provisioned with 32 core CPU, 64GB RAM and running Ubuntu18.04 OS. The microservices in our system architecture such as {\it Goods Processor} and {\it Invoice Processor} are deployed on the RedHat OpenShift container platform.
We simulate the global trade ecosystem using a blockchain platform. We map each ecosystem participant to a membership service provider (MSP) in the blockchain network. Each MSP consists of one or more peers of the blockchain network. Each peer is deployed as a docker container running Hyperledger Fabric version $1.4.1$ on a dedicated SoftLayer ~\cite{softLayer} VM provisioned with 32 core CPU, 64GB RAM and running Ubuntu18.04 OS. The microservices in our system architecture such as {\it Goods Processor} and {\it Invoice Processor} are deployed on the RedHat OpenShift container platform.

We are currently working on the proposed system with couple of anchor clients (companies in consumer goods and retail industries) who have shared anonymized data from their live environments. This data comprises of more than $440 K$ anonymized EDI documents (POs, DAs, RAs, and CIs), which is used to analyze the performance and scalability of our system. In our simulation, each call to the {\it computeCA} transaction generates the CA for a $\langle PO, DA, RA, CI \rangle$ tuple selected at random from the pool of these $440K$ EDI documents.  Subsequently, each call to the {\it computePAs} transaction uses the computed CA to generate the PAs for the associated supplier and carriers. We compute the CA under all the {\it four} claim categories together in our experiments, since evaluating permance meterics for both the passes of CA computation together gives us the minimum performance under each pass (i.e., {\it Pass-1} and {\it Pass-2} in Section \ref{sec:caComputation}) independently. And the {\it computePAs} transaction generates PA for four organizations (supplier, OCM, either origin land carrier or destination land carrier, and ocean carrier) as per the computation logic in Section \ref{sec:paComputation}. For both these transactions {\it computeCA} and {\it computePAs}, we have collected two important metrics: {\it latency} and {\it throughput}. This has been realized with the help of Hyperledger Caliper~\cite{caliper}. Hyperledger Caliper is a blockchain benchmarking tool that allows users to generate performance analysis reports for their blockchain-based implementations. The transactions {\it computeCA} and {\it computePAs} represent different smart contracts and are implemented as chaincode written in Go language. These smart contracts are deployed on all the network peers. The transaction {\it computePAs} is lightweight in comparison to {\it computeCA}. The same is observed in all our experiments, where the transaction {\it computeCA} consistently exhibit lower throughput (and higher average latency) than the transaction {\it computePAs}.

Hyperledger Fabric configuration parameters can be tuned appropriately to achieve optimum performance numbers for a blockchain solution ~\cite{thakkar2018performance}. However, in this paper, we have used the default values for all the configuration parameters of Hyperledger Fabric except that the {\it block size} is set to $100$ transactions and the {\it block timeout} is set to $500ms$.  We have run each experiment $10$ times and submitted a total of $100K$ transactions in each run. We report the experimental results averaged across all these runs.

\subsection{Increase in transaction send rate vs. performance}
\label{sec:evaluationTPS}
\begin{figure}[ht]
    \centering
    \includegraphics[width=1.0\columnwidth]{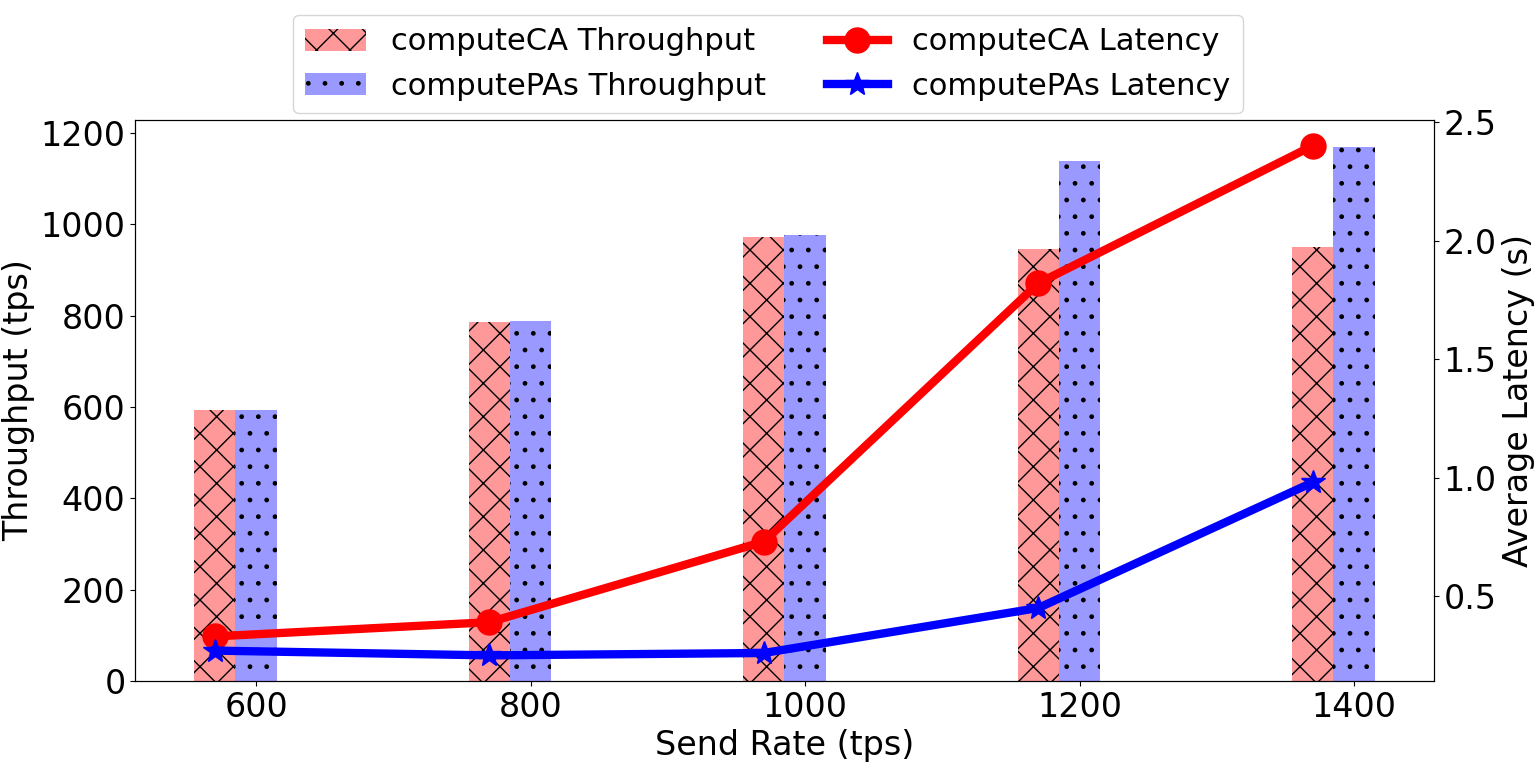}
    \caption{Effect of increasing send rate on performance}
    \label{graph:tpsa}
\end{figure}

%\begin{figure}[ht]
%    \centering
%   \includegraphics[width=0.8\columnwidth]{graphs/plots/4peer4orgSameDCIncTPSLatency.png}
%    \caption{Effect of increasing send rate on latency}
%    \label{graph:tpsb}
%\end{figure}

Figure \ref{graph:tpsa} depicts the impact of increasing transaction send rate on the throughput and latency of {\it computeCA} and {\it computePAs} transactions. It was observed that with the increasing transaction send rate, the {\it latency} for both the transactions increase.  Also, the transaction {\it throughput} initially increases and then saturates around $960$ tps for {\it computeCA} transaction and $\approx1170$ tps for {\it computePAs} transaction.

Note that, this experiment depicts a hypothetical scenario with exorbitantly high transaction rates. Currently, the global trade industry handles $\approx802$ million TEUs (twenty-foot equivalent units) per year \cite{currentGlobalTPS}, which translates to $\approx25$ TEUs per second. Therefore, we conclude that the proposed system easily supports the shipment throughput in the current global trade ecosystem \cite{currentGlobalTPS} and will scale well even for future workloads.

\subsection{Increase in number of peers vs. performance}
\begin{figure}[ht]
    \centering
    \includegraphics[width=1.0\columnwidth]{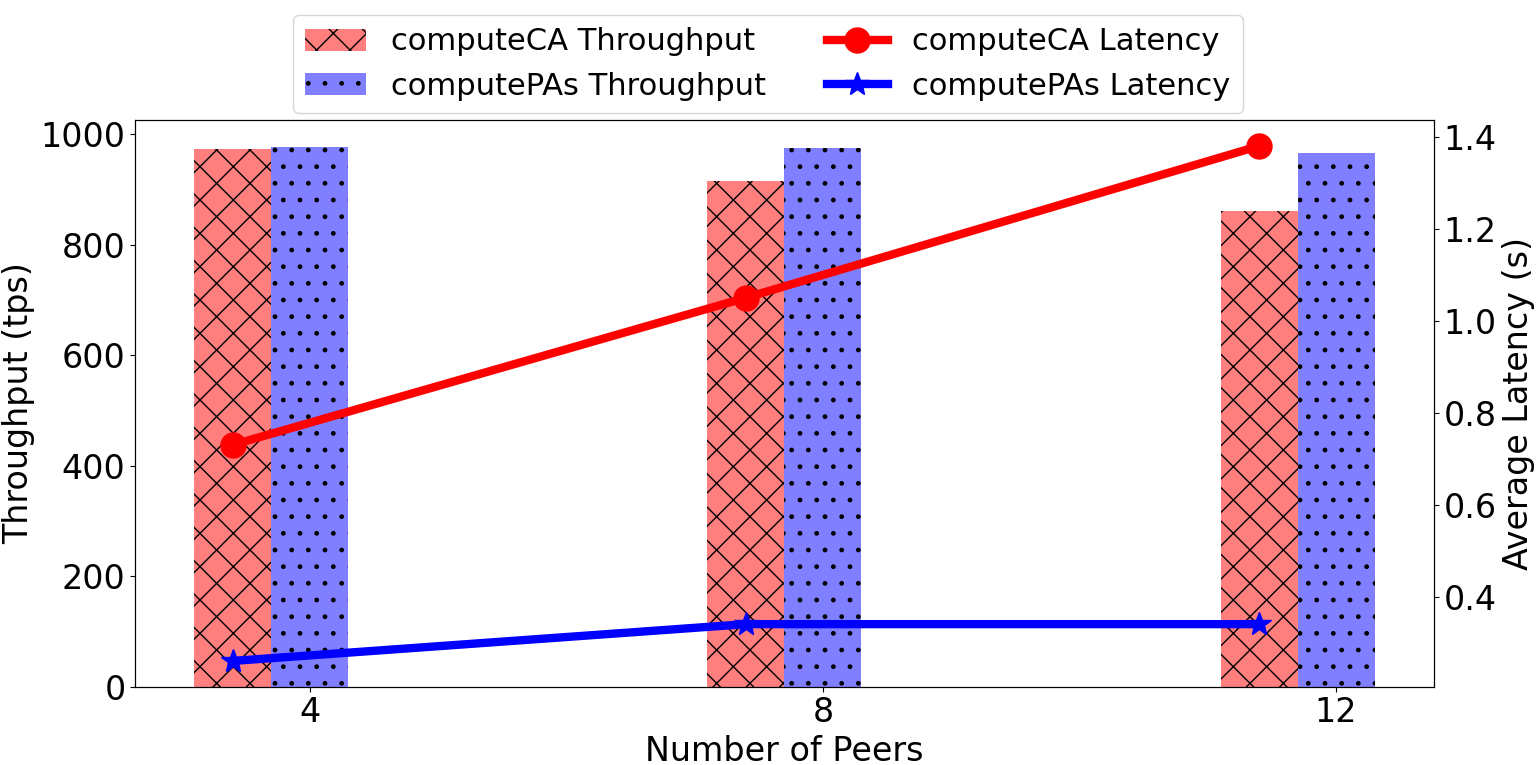}
    \caption{Effect of increasing number of peers on performance}
    \label{graph:noofpeersa}
\end{figure}

%\begin{figure}[ht]
%    \centering
%    \includegraphics[width=0.8\columnwidth]{graphs/plots/IncPeer4orgSameDC1000TPSLatency.png}
%    \caption{Effect of increasing number of peers on latency}
%    \label{graph:noofpeersb}
%\end{figure}

In this experiment, we simulate the increase in the number of participants in the goods trade network by increasing the number of peers in our blockchain network. The transaction send-rate is configured at $1000$ tps, and its impact on {\it throughput} and {\it latency} for the transactions {\it computeCA} and {\it computePAs} is depicted in figure \ref{graph:noofpeersa}. The design of Hyperledger Fabric requires each transaction to collect endorsements from a set of peers in the blockchain network. The number of such peers is governed by the endorsement policy \cite{endorsementPolicy} for the network. Typically, to provide reliability and fault tolerance, the endorsement policy states that a majority of peers should endorse a transaction. Therefore, as the number of peers increases, the number of endorsements collected and validated also increases. This causes transaction latency to increase and throughput to decrease. We observe a similar phenomenon even in our system.

The increase in the number of network peers results in increase of {\it latency} for {\it computeCA} transaction but it has little to no impact on {\it latency} of {\it computePAs} transaction.  Also, the {\it throughput} for {\it computeCA} transaction decreases slightly with the increase in number of peers but we see almost no impact in the case of {\it computePAs} transaction. The reason for this trend is that the transaction send rate ($1000$ tps)  is above the maximum {\it throughput} of {\it computeCA} transaction but well below the maximum {\it throughput} of {\it computePAs} transaction (refer Section \ref{sec:evaluationTPS}).

\subsection{Geographic distribution of peers vs. performance}
\begin{table}
\centering
\caption{Throughput and tatency for 1 DC vs 5 DCs deployments}
    \label{tab:diffDC}
\begin{tabular}{|l|l|l|l|l|}
\hline
\multicolumn{1}{|c|}{\multirow{2}{*}{\begin{tabular}[c]{@{}c@{}}\textbf{Transaction}\\ \textbf{Type}\end{tabular}}} & \multicolumn{2}{l|}{\textbf{Throughput(tps)}} & \multicolumn{2}{l|}{\textbf{Average Latency(s)}} \\ \cline{2-5} 
\multicolumn{1}{|c|}{}                                                                            & 1 DC           & 5 DCs           & 1 DC           & 5 DCs            \\ \hline
computeCA                                                                                    & 973.4          & 238.7         &0.73           & 18.87         \\ \hline
computePAs                                                                             & 977.0          & 323.0         &0.26          & 12.37             \\ \hline
\end{tabular}
\end{table}

Table \ref{tab:diffDC} depicts the impact on throughput and latency as the peers get more distributed over different data centers (DCs). This experiment simulates the scenario where the participants of the global trade network belong to diverse geographies. We considered the setting where blockchain peers were spread across four geographically different data centers and the orderer node was located in a fifth geographic cloud data center. We observe an increase in {\it latency} and decrease in {\it throughput} for both transactions as peers get widely distributed over geography (i.e., as participants from different geographic locations onboard the global trade network).

It was observed that when blockchain nodes were located in the same geographical data center, the ping latency among the nodes was $< 1 ms$, whereas when the nodes were spread across $5$ data centers, the ping latency among the nodes varied between $[50, 130] ms$. Hence, we infer that network latency is the reason behind degradation in performance. The performance in such scenarios can be improved by using a low latency network. However, it is important to note that even without such an optimization, the proposed system can easily support $>10$ times the load of the current global shipping industry\cite{currentGlobalTPS}.
\section{Conclusion}
\label{sec:conclusion}
In this paper, we discussed the need for a blockchain-based accounts payable system that eliminates the process redundancies (accounts payable vs. accounts receivable), enables efficient invoice processing, and reduces the amount of time spent in reconciling disputes between the transacting participants in goods trade (domestic and global). We provided details of our blockchain-based accounts payable system supporting trusted invoice processing and transparent dispute resolution. The claims against a PO are generated under four different categories and aggregated to produce the claim advice. The payment advices for supplier and carriers are generated by using the invoices for supplier and carriers respectively together with the claim advice generated by our accounts payable system and real-time events from TradeLens. The computed CA and PAs go through the reconciliation process before the payment gets processed by the shipper. We have realized the system proposed in this paper using Hyperledger Fabric as the underlying blockchain platform (however any other permissioned blockchain platform can also be used here) and a cloud microservices architecture. We showcased the performance of different smart contract modules (i.e., transactions) supported by our system using a representative goods trade ecosystem. The results show that its practical to deploy one such system in real-world customer environments. We are currently experimenting with the use of the proposed system in collaboration with the participants of the TradeLens platform.
 
 % Our system also exposes a comprehensive set of APIs to offer a variety of functionality to the goods trade participants, namely: (i) to import the various EDI documents PO, DA, RA, supplier invoice (CI), and carrier invoices in GS1 standard (ii) to provide role-based access to the shared data on the blockchain (iii) to export the generated CA and PAs into the ERP systems of the different organizations for downstream processing (e.g., payment processing).

\bibliographystyle{IEEEtran}
\bibliography{icbc}

\end{document}